\documentclass[11pt,a4paper]{article}
\usepackage[utf8]{inputenc}
\usepackage[T1]{fontenc}
\usepackage{lmodern}
\usepackage{microtype}
\usepackage[margin=2.5cm]{geometry}
\usepackage{setspace}
\usepackage[skip=0mm,indent=14pt]{parskip}
\usepackage{graphicx}
\usepackage{booktabs}
\usepackage{enumitem}
\usepackage{hyperref}
\hypersetup{colorlinks=true,linkcolor=blue,citecolor=blue,urlcolor=blue}
\usepackage{titling}
\usepackage{fancyhdr}
\usepackage{caption}
\usepackage{amsmath,amssymb}
\usepackage{aas_macros}

\usepackage[super,sort&compress]{natbib}
\usepackage[flushleft]{paralist}

\usepackage[printwatermark]{xwatermark}
\usepackage{xcolor}
\usepackage{graphicx}

\pretitle{\vspace*{-1cm}\begin{center}\LARGE\bfseries}
\posttitle{\par\end{center}\vskip 0.5em}
\preauthor{\begin{center}\large}
\postauthor{\end{center}}
\predate{\begin{center}\large}
\postdate{\end{center}}

\begin{document}

\begin{titlepage}
  \centering
  {\Huge\bfseries Expanding Horizons \\[6pt] \Large Transforming Astronomy in the 2040s \par}
  \vspace{1.0cm}

  {\LARGE \textbf{Dynamical binary interactions in the 2040s}\par}
  \vspace{1cm}

  \begin{tabular}{p{4.5cm}p{10cm}}
    \textbf{Scientific Categories:} & stars: novae, cataclysmic variables, stars: supernovae, stars: binaries, stars: mass loss
 \\
    \\
    \textbf{Submitting Authors:} & \textbf{Nadejda Blagorodnova} (University of Barcelona, Spain), \\
    & nblago@fqa.ub.edu \\
    & \textbf{Ond\v{r}ej Pejcha} (Charles University, Czechia),\\
    & ondrej.pejcha@matfyz.cuni.cz \\
    & \textbf{Tomek Kami\'{n}ski} (Nicolaus Copernicus Astronomical Center, Polish Academy of Sciences), tomkam@ncac.torun.pl\\

    \\
    \textbf{Contributing authors:} & \textbf{Yongzhi Cai} (Yunnan Observatories, Chinese Academy of Sciences), \textbf{Kishalay De} (Columbia University), \textbf{Nancy Elias-Rosa} (INAF Padova), \textbf{Jim Fuller} (Caltech), \textbf{Hongwei Ge} (Yunnan Observatories, Chinese Academy of Sciences), \textbf{David Jones} (Instituto de Astrof\'{i}sica de Canarias), \textbf{Stephen Justham} (MPA Garching), \textbf{Viraj Karambelkar} (Columbia University), \textbf{Jakub Klencki} (MPA Garching), \textbf{Elena Mason} (INAF -- OATS), \textbf{Brian Metzger} (Columbia University), \textbf{Andrea Pastorello} (INAF Padova), \textbf{Andrea Reguitti} (INAF Padova), \textbf{Friedrich R\"{o}pke} (Heidelberg Institute for Theoretical Studies), \textbf{Steven Shore} (Pisa University), \textbf{Giorgio Valerin} (INAF Padova) \\

  \end{tabular}

  \vspace{1cm}

  \textbf{Abstract:}

  \vspace{0.5em}
  \begin{minipage}{0.9\textwidth}
    \small
    Dynamical binary interactions such as common envelope (CE) evolution or stellar mergers are a critical phase in the formation of a wide variety of binary phenomena, ranging from blue stragglers to type I supernovae (of all flavours, a, b and c), $\gamma$-ray bursts to bipolar planetary nebulae, Thorne-Zytkow objects to X-ray binaries. In 2040s, the urgency of resolving long-standing questions regarding the physics behind the dynamical interaction stages and the absolute and relative frequencies of binary evolutionary pathways will only increase owing to rapidly expanding population statistics of gravitational wave events. Here, we argue that multi-wavelength observations (spectroscopy and photometry), linear spectropolarimetry, and interferometry of a large number of Luminous Red Novae, a particular class of transients associated with dynamical binary interactions, will provide unprecedented details about the underlying interaction physics. A breakthrough will be achieved by a tenfold or larger increase in identifications of transient-type events from interacting binaries and their follow-up with instrumentation that provides at least 10 times better angular resolution, 100 times better spectral resolution, and $\sim$100 times higher sensitivity than 2030s facilities.  \end{minipage}

\end{titlepage}


\section{Introduction and Background}
\label{sec:intro}

\begin{figure}
\centering
\includegraphics[width=0.855\textwidth]{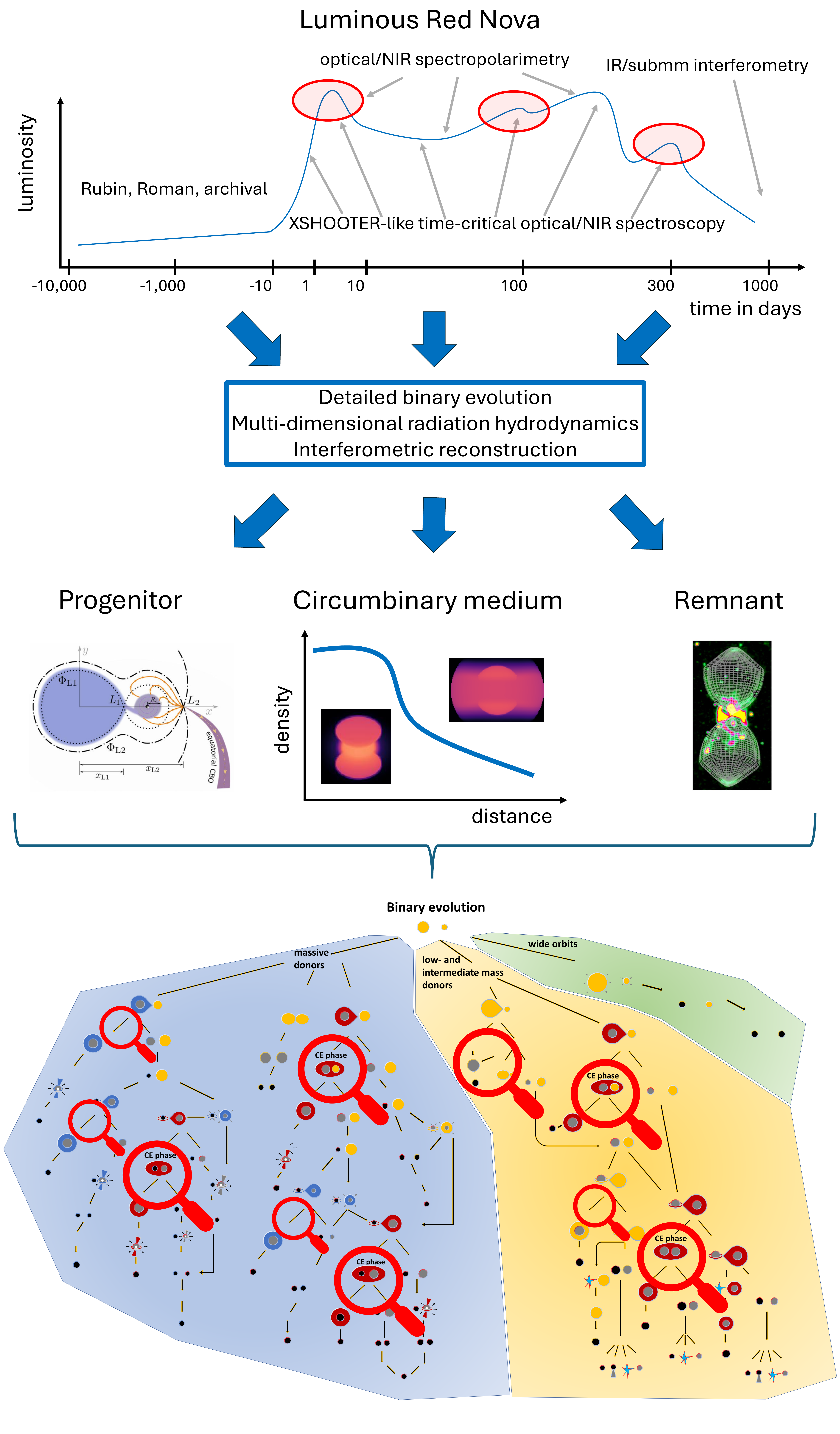}
\caption{From LRNe observations to constraints on diverse binary evolution pathways. The elements of the plot were adapted from the literature\cite{han20,kaminski_ckvul,lu23,kirilov25}. The magnifying glass shows the phases of unstable mass transfer leading to binary mergers, and common envelope evolution. \label{fig:plan}}
\end{figure}

Recent studies have shown that stellar interactions are an intrinsic part of stellar evolution, especially at the most massive end, where more than 70\% of stars have a closeby companion \citep{Sana2012Sci}. Interactions between binary companions through mass transfer can significantly alter their evolutionary pathways, leading to a wide range of astrophysical phenomena, such as thermonuclear and stripped core-collapse supernovae, stripped hot stars, X-ray binaries, novae, and compact binary sources, which are the progenitors of gravitational wave sources. The lower panel of Fig.~\ref{fig:plan} illustrates the comprehensive tree of possible binary evolutionary pathways. Given the fast growth of multi-messenger, multi-wavelength astronomy, the 2040s will be a promising decade for jointly exploring this multi-dimensional information on binary evolution, combining detailed studies of individual objects to understand the particular physics at play with large population studies, where the statistics of different multi-channel populations (gravitational wave sources, optical and infrared transients) will be combined to obtain the most detailed picture of binary stellar evolution.

In the past twelve years, a particular type of astrophysical transient has emerged as a promising tool for the study of dynamical binary interactions. These transients, called Luminous Red Novae (LRNe), represent the final, most dramatic interaction between two stars, where an extended period of runaway mass transfer from an expanding donor star leads the system to enter the Common Envelope (CE) phase \citep{Paczynski1976IAUS}, where both stars orbit inside the donor's envelope. This stage concludes with either a partial or a complete ejection of the envelope. While partial ejections will lead to mergers, full CE ejections represent a promising channel for the formation of gravitational wave sources from the isolated binary channel.

As shown by the magnifying glass icon in the lower panel of Fig.~\ref{fig:plan}, nearly all possible branches of binary evolution necessarily involve at least one CE stage. A thorough characterization of this stage is vital for constraining both the outcomes and the intrinsic rates of binary-evolution processes. This includes a zoo of interaction products that not only were dynamically affected by the interaction, but that also changed their internal structure and surface chemical composition (e.g., extreme helium stars, H-deficient carbon stars, J-type carbon stars).

Advances in the field of CE evolution and binary evolution rely on four main actions: 1) Develop detailed 2D/3D radiation-magneto-hydrodynamical simulations capturing the most important physics and translate them into observable signatures. 2) Obtain high-quality observational data that can verify the model predictions and constrain the physical processes to input into the detailed modeling. 3) Feed the improved and tested physics into large binary population synthesis models. 4) Calibrate these models using population statistics and rates to make inferences about additional binary channels. 

The study of the particular physics involved in each event involves several stages: understanding the unique properties of the binary star undergoing dynamical interaction; understanding the mass-transfer process; understanding the dynamical ejection; and understanding the remnant relaxation and subsequent evolution. Fortunately, LRNe present an ideal laboratory in which all these stages occur over only a few years, yielding a uniquely complete, beginning-to-end perspective on the evolution of the underlying binary interactions (top panel of Fig.~\ref{fig:plan}). So far, progress has been made in this field through the study and modeling of LRNe, together with the archival progenitor binaries and interferometric observations of the remnants in several pathfinder studies \cite{blagorodnova21,Cai2022AA,kaminski21,mason22,steinmetz24}. 

The 2040s will open up a vastly larger range of LRNe timescales, luminosities, and energies, revealing phenomena well beyond what is known today. With the completion of the Euclid, Nance Grace Roman, and Vera Rubin's large sky surveys, archival data for thousands of LRNe progenitors as well as the precursor evolution leading to the mergers in nearby galaxies will be available, allowing us to identify the quiescent systems associated with almost all close-by events. The operations of JWST would also have allowed us to compile mid-IR data on the dust formation processes and late-time evolution of the remnants. However, the detailed study of the pre-dynamical and dynamical ejection process and the remnant will continue to be challenging. Due to the faint intrinsic luminosity and slow ejection velocities ($5-1000$\,km/s) in these events, the most informative observations require fast response, a high S/N ratio, and a high spectral and spatial resolution, and should provide information on the dynamics, composition, and geometry of the ejecta. Even with a large facility such as ELT starting operations in the 2030s, only a handful of objects would have received the necessary observations. IR and mm interferometric observations of the remnant are limited by the baseline and sensitivity of existing facilities. 
Because of their high diversity in light curve shapes and spectroscopic signatures, detailed physical tests and inferences about the broader population will not be possible with current and planned facilities. 
Next-generation instrumentation must deliver transformative capabilities: tenfold improved sensitivity, as well as hundred times better spectroscopic and angular resolution to study dozens of individual transients from binary interactions per year in exquisite detail, while simultaneously enabling coarser monitoring for their complete population (including the faintest events) out to 150\,Mpc, representing 10$^3$-10$^4$ events per year.

\section{Open Science Questions in the 2040s}
\begin{compactenum}
    \item \textbf{What happens at each interaction stage?} Relative to our Fig.~\ref{fig:plan}, we need to answer what physics are involved in each magnifying glass step: what are the processes leading to CE? What impact does the evolutionary phase of the donor and/or accretor have on the CE process and outcome? How is mass ejected? Did the binary survive? How were the internal structure and surface composition affected? What processes control the evolution of the remnant?
    \item \textbf{What are the absolute and relative frequencies of binary evolutionary branches?} Relative to our Fig.~\ref{fig:plan}, answering this question will require combining detailed studies of the LRNe rate with progenitor information, understanding of individual interaction stages, and population studies of the progeny and surviving binaries. 
\end{compactenum}

\section{Technology and Data Handling Requirements}
\label{sec:tech}

\begin{compactitem}
    \item \textbf{Fast photometric follow-up} ($<$8h of discovery) is critical to uncover early shock interaction. Up to year-long  (or up to a liming mag of $r$=27) multi-wavelength UV, optical, and IR photometry lightcurves are needed with sufficient cadence ($\leq$3 days) to determine the event energetics.  
    \item \textbf{Fast spectroscopic follow-up} ($<$8h of discovery) to perform LRN flash spectroscopy to detect low-velocity CSM features (a few km/s), later erased by the expanding ejecta. This requires North and South Hemisphere coverage, and spread over different time zones.
    \item \textbf{High resolution optical and NIR spectroscopy} (R$\sim$40,000) for classification and follow-up of new LRNe. The data should provide S/N$>$40 (in the continuum) for a 20\,mag object with $<$1\,h exposure time.
    \item Follow-up \textbf{optical-NIR linear spectropolarimetry} of R$>$10000 should provide S/N$>$150 for a 20 mag object to distinguish line polarization vs. continuum and reliably detect polarization changes as the transient evolves.
    \item \textbf{Mid-infrared and mm/submm photometry (up to mas resolution) and spectra} to study warm dust in interacting and post-interaction systems within 20\,Mpc, crucial for determining true bolometric luminosities and dust-formation rates.
\end{compactitem}
\vskip 1mm



\bibliography{refs}  

\end{document}